%% file: conference_101719.tex
\documentclass[sigconf]{acmart}
\copyrightyear{2024}
\acmYear{2024}
\setcopyright{rightsretained}
\acmConference[AST '24]{5th ACM/IEEE International Conference on Automation of Software Test (AST 2024)}{April 15--16, 2024}{Lisbon, Portugal} \acmBooktitle{5th ACM/IEEE International Conference on Automation of Software Test (AST 2024) (AST '24), April 15--16, 2024, Lisbon, Portugal}\acmDOI{10.1145/3644032.3644463}
\acmISBN{979-8-4007-0588-5/24/04}

\acmConference[AST 2024]{5th International Conference on Automation of Software Test}{April 2024}{Lisbon, Portugal}

\usepackage{algorithmic}
\usepackage{graphicx}
\usepackage{textcomp}
\usepackage{xcolor}
\def\BibTeX{{\rm B\kern-.05em{\sc i\kern-.025em b}\kern-.08em
    T\kern-.1667em\lower.7ex\hbox{E}\kern-.125emX}}

\usepackage[nameinlink]{cleveref}

\usepackage{blkarray} %
\usepackage[inline,shortlabels]{enumitem} %
\usepackage{mdframed} %
\usepackage{float} %
\usepackage{booktabs} %
\usepackage{caption}

\usepackage{booktabs}
\usepackage{tabularx}
\usepackage{multirow}

\usepackage{array}
\usepackage{xspace}
\usepackage{colortbl}
\usepackage{xcolor}

\usepackage{lscape} %

\newcolumntype{Y}{>{\centering\arraybackslash}X}

\usepackage{comment}

\usepackage{tablefootnote}

\newcommand{\bugs}{19\xspace}
\newcommand{\apps}{17\xspace}

\usepackage{xfp}

\newcommand{\repoTool}{\url{https://gitlab.com/sal-unimib-anonymization/anonymization-android-tool}\xspace}
\newcommand{\repoExperiments}{\url{https://gitlab.com/sal-unimib-anonymization/experimentation}\xspace}

\newcolumntype{s}{>{\hsize=.5\hsize}X}
\newcolumntype{m}{>{\hsize=.7\hsize}X}

\makeatletter
\newcommand{\rqitem}[2][]{%
  \ifblank{#1}{%
  \item #2%
  }{%
  \item[#1] #2%
  }%
  \protected@edef\@currentlabelname{#2 (\theenumi)} 
}
\makeatother

\makeatletter
\newcommand{\linebreakand}{%
  \end{@IEEEauthorhalign}
  \hfill\mbox{}\par
  \mbox{}\hfill\begin{@IEEEauthorhalign}
}
\makeatother

\usepackage[]{xcolor}

\begin{document}

\title{Anonymizing Test Data in Android: Does It Hurt?
}

\author{Elena Masserini}
\email{e.masserini2@campus.unimib.it}
\affiliation{%
  \institution{University of Milano - Bicocca}
  \city{Milan}
  \country{Italy}
}

\author{Davide Ginelli}
\email{davide.ginelli@unimib.it}
\affiliation{%
  \institution{University of Milano - Bicocca}
  \city{Milan}
  \country{Italy}
}

\author{Daniela Micucci}
\email{daniela.micucci@unimib.it}
\affiliation{%
  \institution{University of Milano - Bicocca}
  \city{Milan}
  \country{Italy}
}

\author{Daniela Briola}
\email{daniela.briola@unimib.it}
\affiliation{%
  \institution{University of Milano - Bicocca}
  \city{Milan}
  \country{Italy}
}

\author{Leonardo Mariani}
\email{leonardo.mariani@unimib.it}
\affiliation{%
  \institution{University of Milano - Bicocca}
  \city{Milan}
  \country{Italy}
}

\begin{abstract}
Failure data collected from the field (e.g., failure traces, bug reports, and memory dumps) represent an invaluable source of information for developers who need to reproduce and analyze failures. Unfortunately, field data may include sensitive information and thus cannot be collected indiscriminately. Privacy-preserving techniques can address this problem anonymizing data and reducing the risk of disclosing personal information. However, collecting anonymized information may harm reproducibility, that is, the anonymized data may not allow the reproduction of a failure observed in the field. 
In this paper, we present an empirical investigation about the impact of privacy-preserving techniques on the reproducibility of failures. In particular, we study how five privacy-preserving techniques may impact reproducibilty for \bugs bugs in \apps Android applications. Results provide insights on how to select and configure privacy-preserving techniques.
\end{abstract}



\maketitle

\section{Introduction}
\label{sec:introduction}

Collecting bug reports and information about the failures experienced by end-users while interacting with their applications is extremely important to reveal bugs~\cite{DBLP:journals/tse/ZimmermannPBJSW10,DBLP:journals/tse/ZouLCXFX20}, and improve the quality and the reliability of the applications. Indeed, several problems are detected only once the software has been released~\cite{Gazzola:FieldFail:ISSRE:2012}, and the extensive collection of failure data is a key factor to enable the reproduction of the bugs, and later their correction.

Several approaches have been defined to reproduce failures from runtime data extracted from the field. For instance, failures have been reproduced starting from the flow of events executed in the app immediately before a crash~\cite{DBLP:conf/qrs/SunYLZX19}, from executions traces with the operations performed before a crash~\cite{DBLP:conf/wcre/NayrollesHTL15,DBLP:conf/icse/JinO12,DBLP:journals/tse/ChenK15}, as well as from the content of the stack trace~\cite{DBLP:journals/tse/SoltaniPD20,DBLP:conf/iwpc/WhiteVJBP15}, and bug reports~\cite{DBLP:conf/icsr/ZhaoMYZP19,DBLP:conf/icse/ZhaoYSLZZH19}. Despite the benefit of collecting data from the field to reproduce failures, user data can be fairly collected only by taking the sensitivity of the data into consideration. Indiscriminately collecting data may reveal sensitive information that should not be available outside the boundary of the app. For instance, failure traces may include sensitive information such as age, gender, financial data, and personal interests.

In specific cases, data can be partially anonymized. For instance,  concerning the data stored in databases, \textit{kb}-Anonymity can be used to mitigate the issue of sharing sensitive information through the databases used for testing~\cite{DBLP:conf/pldi/BudiLJL11}. When the execution path of the failure is available and symbolic execution is applicable to the target program, new synthetic executions that reproduce the failures might be derived to also mitigate issues with sensitive data~\cite{DBLP:conf/issre/MatosGR15,DBLP:conf/edcc/LouroGR12,DBLP:conf/asplos/CastroCM08}. 

The field of data mining has been investigating this challenge for several years defining a number of privacy-preserving techniques that can be used to alleviate the problem of incidentally disclosing sensitive information~\cite{DBLP:journals/csur/FungWCY10, DBLP:journals/access/XuJWYR14, DBLP:conf/bigdatasec/MurthyBRR19}. These techniques work by applying generalization or suppression operations to the data, so that the original information is not immediately available anymore~\cite{DBLP:journals/access/MendesV17}. For instance, a string \textsc{123456} representing an account number could be automatically rewritten as a random string of the same length, such as \textsc{xhfprt}. These techniques can be readily applied to the data collected from the field to prevent disclosing sensitive data to third-parties.

While privacy-preserving techniques can clearly eliminate, or reduce, privacy issues, their impact on the capability of revealing failures has not been studied so far. Indeed, using anonymized data to reproduce failures is harder than using clear text data. That is, \emph{protecting the privacy of the users and facilitating the reproduction of failures experienced in the field are two competing goals}.

In this paper we propose the first, to the best of our knowledge, empirical study about the impact of privacy-preserving techniques on failure reproduction. We focused our study on failures experienced by users of mobile apps due to the popularity of mobile applications and their exposure to privacy issues~\cite{PrivacyMobApp:IST:2021}. Our study considers \bugs bugs in \apps open source Android applications, and discloses insights about the trade-off between guaranteeing the privacy of the users and easing the reproduction of failures.
In particular, we show that there is no unique choice about the privacy-preserving techniques to be used. Different contexts may require different techniques depending on the aspect to privilege.

This paper is organized as follows. \Cref{sec:background} introduces and rigorously defines privacy-preserving techniques. \Cref{sec:experiment-design} describes the design of the experiment conducted to evaluate the impact of privacy-preserving techniques on failure reproduction. \Cref{sec:results} reports the empirical results and answers our research questions. \Cref{sec:related-work} discusses related work. Finally, \Cref{sec:conclusion} provides final remarks.

\section{Privacy-Preserving Techniques}
\label{sec:background}

Privacy-preserving techniques can be used to effectively anonymize data. This section defines the techniques that we considered in our study, describing how we adapted them to the problem of failure reproduction, when necessary. 

Privacy-preserving techniques are typically used in the context of data mining, especially with records of databases. In fact, data contained in tables do not usually satisfy privacy requirements, and thus they cannot be shared without applying anonymization operations~\cite{DBLP:journals/csur/FungWCY10}. These operations may target individual records or sets of records. The former class of operations is useful when a third-party that accesses the data can essentially access only to individual records, and cannot compare the (anonymized) records between them. The latter class of operations is useful when a third-party can access the full set of records, and thus may infer facts by cross-analyzing the content of multiple (anonymized) records. In such a case, privacy-preserving techniques must consider the full set of records when anonymizing the individual records to prevent the incidental disclosure of sensitive information.

In the case of software failures, they are usually experienced once a while for the released apps, and only few failures are normally collected from a same user. To guarantee that user data is fairly collected, the application of strategies specifically designed to deal with large sets of repeated failures collected from the same users is likely not needed. For this reason, we focus on privacy-preserving techniques that can be applied to individual records.

In this paper, we consider failure traces consisting of streams of GUI events executed on the app when the failure occurred, as done in many failure reproduction techniques, such as CaRCrash~\cite{DBLP:conf/qrs/SunYLZX19} and ReCDroid~\cite{DBLP:conf/icse/ZhaoYSLZZH19}. That is, a failure trace is a sequence of events $(a_i, w_i, d_i)$ where $a_i$ is a GUI action (e.g., a click event) performed on a widget $w_i$ (e.g., a button) possibly using data $d_i$ (e.g., the text entered into an input field). The set of values $d_i$ that occur in a failure trace are the data values subject to the anonymization process. We do not consider in our experiment the anonymization of other elements, such as the action or the widget. That is, we study how to prevent the failure trace from disclosing information such as the age, the address, or the personal income through the data values $d_i$ entered into a form, while it is out of the scope of the study to hide the fact that a user has registered into an app by clicking on the \textsc{Register} button. 

The operations performed by privacy-preserving techniques to anonymize data vary based on the type of data. In particular, it is possible to distinguish three different classes of data to be analyzed: continuous values,  categorical values, and string values. Continuous values are numeric values (e.g., someone's age or income) that can be used, for instance, as part of arithmetic operations. Categorical values are enumerated values that cannot be normally used as part of arithmetic operations~\cite{DBLP:series/ads/Domingo-Ferrer08}. Finally, string values are sequences of alpha-numeric characters.

We now present the privacy-preserving techniques based on the type of anonymization strategy they implement: generalization, suppression, and perturbation. %
\autoref{tab:privacy-preserving-techniques} shows the specific techniques that we considered (Column \emph{Technique}), classified according to the strategy they implement (Column \emph{Strategy}) and associated with type of data that they can be applied to (Columns \emph{Continuous}, \emph{Categorical}, and \emph{String}).
The set of selected techniques reflects the taxonomy proposed by Mendes et al.~\cite{DBLP:journals/access/MendesV17}. We have excluded the {\it anatomization strategy} presents in the taxonomy, since it is strictly related to databases and cannot be applied to our context, and both the Top/Bottom Coding and the Post-Randomization (PRAM) techniques since our dataset does not include cases where they can be applied. %

\begin{table}[ht]
\footnotesize
\centering
\caption{Overview of privacy-preserving techniques.}
\label{tab:privacy-preserving-techniques}
\resizebox{\columnwidth}{!}{
\begin{tabularx}{\columnwidth}{p{0.09\textwidth}p{0.14\textwidth}YYY}
\toprule
Strategy & Technique & Continuous & Categorical & String \\
\midrule
\textbf{Generalization}& Global Recoding & \checkmark & \checkmark & - \\
& Top Coding & \checkmark & - & - \\
& Bottom Coding & \checkmark & - & - \\
& Rounding & \checkmark & \checkmark & - \\
\midrule
\textbf{Suppression}& Local Suppression & \checkmark & \checkmark & \checkmark\\
& Special Char Driven LS & - & - & \checkmark \\

\midrule
\textbf{Perturbation}& Noise Addition &  \checkmark & - & - \\
& PRAM & - & \checkmark & - \\
\bottomrule
\end{tabularx}
}
\end{table}

\smallskip
\subsection{Generalization Techniques}
Techniques that belong to this strategy replace values with more general ones~\cite{DBLP:journals/access/MendesV17}. %
These privacy-preserving techniques disclose some general information, while hiding the original value.   

\subsubsection*{\textbf{Global Recoding}}
\paragraph*{\textbf{Definition:}} 
Global Recoding anonymizes a value by only disclosing information about the interval it belongs to. This technique can be applied to both categorical and continuous variables. In the former case, Global Recoding anonymizes a value by combining several categories into fewer ones. For example, if the categorical value represents different age groups (e.g., newborns, infants, toddlers, kids, and adults), Global Recoding may reduce them into two groups (e.g., 'baby' and 'kids or older'). In the latter case, Global Recoding replaces a variable with its interval. For example, the numerical age value can be replaced with its categorical age group~\cite{Templ2017}.
\paragraph*{\textbf{Failure Reproduction:}} 
In the context of failure reproduction, replacing a categorical or continuous value with its interval (e.g., replacing the categorical input \texttt{infant} or the numerical input \texttt{1} with the category \texttt{baby}) would make the failure trace non-executable. In fact, the new value would not be processable by the application that expects either values from a specific enumeration of categories or numerical values. To obtain a processable input, and thus to attempt to reproduce the failure from the anonymized trace, the values anonymized with Global Recoding are then replaced with random concrete values within the anonymized interval. 
\paragraph*{\textbf{Example:}} 
If a variable in the range [0, 10] is anonymized according to the sub-intervals [0, 5) and [5, 10], and the value to anonymize is 4.0, Global Recoding replaces the original value with the interval [0, 5). Failure reproduction shall generate random values within this interval to attempt to reproduce the failure. Similarly, if a categorical value \texttt{newborns} is anonymized with a more general category \texttt{baby} (that includes \texttt{newborns}, \texttt{infants}, and \texttt{toddlers}), test generation shall use values in the set \texttt{newborns}, \texttt{infants}, and \texttt{toddlers} to reproduce the failure.

\subsubsection*{\textbf{Rounding}}
\paragraph*{\textbf{Definition:}}  
This technique identifies several rounding points in the domain and maps the input value to be anonymized to the closest rounding point~\cite{DBLP:series/ads/Domingo-Ferrer08}. These rounding points could be identified by dividing the domain into multiple intervals, then selecting the middle point of each interval as rounding point.
\paragraph*{\textbf{Failure Reproduction:}} 
The anonymized value is an actual domain value and thus failure reproduction simply uses the value readily available in the trace. 
\paragraph*{\textbf{Example:}}  
Given an input in the range (0, 10], the rounding points can be defined as the middle points of the intervals (0, 5) and [5, 10], that is, the values 2.5 and 7.5. Every value to be anonymized is mapped to one of these two values.

\subsection{Suppression Techniques} 
Techniques that belong to this strategy entirely drop the
values to be anonymized, or retain minimal information,
to protect privacy~\cite{DBLP:journals/access/MendesV17}.

\subsubsection*{\textbf{Local Suppression}}
\paragraph*{\textbf{Definition:}}
This technique can be trivially applied to any data type (continuous, categorical, and string), since it replaces the input value with a missing value, whose semantics depends on the context~\cite{Templ2017}. For example, considering a record in a database, the corresponding missing value is \verb|NULL|. In the context of Android applications, Local Suppression simply logs the empty string for any input value. 
\paragraph*{\textbf{Failure Reproduction:}} 
In this case, failure reproduction is left with no information about the original value and thus it can only generate a random value coherent with the domain of the original value. In particular, if the original value is continuous, the technique generates a random value within the allowed range. If the original value is categorical, the technique chooses a random element from the set of possible values. If the original value is a string, the technique generates a string that matches a specific regular expression (in such a case, we consider both the case the new string has a length unrelated to the original string or has a length matching the original string). 
\paragraph*{\textbf{Example:}}
In all the cases, the anonymized value is the empty value. The generation is driven by the full range of values allowed by the input field. For instance, a random number between 0 and 100 could be generated for an input field representing the age of a person.

\subsubsection*{\textbf{Special Char Driven Local Suppression}}

\paragraph*{\textbf{Definition:}}
Since sometimes bugs are triggered by anomalous characters that cannot be parsed or processed correctly, we defined a version of the Local Suppression that only preserves the special characters (defined as any non-alphanumeric character, such as *, !, and ?) contained in the value to anonymize. Special characters usually reveal virtually nothing about the original input, but they might be helpful to reproduce misbehaviors.
\paragraph*{\textbf{Failure Reproduction:}} 
The generation works the same than in Local Suppression, but the special characters in the input value are copied in random places within the generated value.
\paragraph*{\textbf{Example:}}
Given  the value \textit{example!} to be anonymized, the technique generates a new random string that includes the special character \texttt{!}, such as \texttt{HQb!Ha}.

\subsection{Perturbation Techniques}
Techniques that belong to this strategy replace the original
values with synthetic values close to the original ones~\cite{DBLP:journals/access/MendesV17, DBLP:journals/csur/FungWCY10}.

\subsubsection*{\textbf{Noise Addition}}
\paragraph*{\textbf{Definition:}}
This technique is typically applied to continuous variables (i.e., to numbers). The general idea is to change the original value by adding or multiplying a stochastic or randomized number (i.e., the noise) to the original data~\cite{Templ2017}.
Given a domain range $[min, max]$, a percentage of noise amplitude $n$, and a value to anonymize $v$, the technique generates a random value in the interval  $[v - n * ( v - min), v + n * (max-v)]$. %
\paragraph*{\textbf{Failure Reproduction:}} 
The anonymized value is an actual domain value and thus failure reproduction simply uses the value readily available in the trace. 
\paragraph*{\textbf{Example:}} 
Given a range [0, 10], a noise 0.30, and the value to anonymize 8.0, Noise Addition generates a random value in the interval $[8.0 - 0.30 * (8.0 - 0) , 8.0 + 0.30 * (10.0 -8.0)] = [5.6, 8.6]$.

\begin{table*}[!ht]
\caption{Reproducible Android application faults}
\label{tab:app-description}
\setlength{\tabcolsep}{4pt}
\resizebox{\textwidth}{!}{
\begin{tabular}{lllp{5in}}
\toprule
\textbf{Application} & \textbf{Domain} & \textbf{Version} & \textbf{Fault description}
\\ \midrule
Binary Eye & Barcode scanner & 1.56.2 & Generating QR codes from some string values cause the app to crash. \\ 
Birday & Birthdays and events & 1.9.0 & The app crashes if the user adds a birthday on February 29th. \\ 
Catima Loyalty & Card management & 2.16.0 & Although only the initial of a card name should be shown in the icon, certain combination of initial characters are all erroneously shown in the icon. \\ 
Catima Loyalty & Card management & 2.8.0 & Expiry date for cards set before 1970 are not shown. \\ 
Contact Diary & Event and contact tracker & 1.2.0 & The app crashes when the input includes a malformed event duration that does not match the pattern hh:mm, such as \texttt{:mm, hh:}. \\ 
Debitum & Debts and lents tracker & 1.4.0 & Transactions are sometimes saved with an amount that slightly differs from the entered one. \\ 
Did I take my meds? & Medicine tracker & 1.6.2 & Some combinations of system time and edited time cause the edited time to be saved as P.M. even if it was A.M., and vice versa. \\ 
EinkBro & Browser & 8.21.0 & Some queries are considered as URLs and lead to webpage not found error. \\ 
Food Scale Droid & Grocery management & 1.2 & App crashes when weights contain a comma. \\ 
GrowTracker & Gardening & 2.5.1 & Entering a value with a dot in \textit{from date} or \textit{to date} fields causes the app crash. \\ 
Money Wallet & Accounting & 4.0.4.1 & Initial amounts in wallets can be incorrectly saved. \\ 
NoNonsense Notes & Notes & 5.5.1 & Closing and re-opening the app after a SD synchronization causes the loss of all notes contained in lists whose name includes a / . \\ 
Simple Calendar & Calendar & 6.19.0 & When importing birthdays from Simple Contacts, the age shown for contacts born before 1970 is incorrect. \\ 
Simple Money Tracker & Accounting & 0.8.9 & If the transaction amount is too big, the app crashes. \\ 
SplitBills & Shared expenses & 0.3.10 & Group names containing a / as a middle or final symbol are not correctly exported. \\ 
Tasks & Habit tracker & 9.7.3 & When creating a task, subtask names containing the sequence \texttt{' @'} are truncated.\\ 
To don't & Negative habits tracker & 1.1 & App crashes - changes are not saved when editing a task with an apostrophe in the name. \\ 
Track \& Graph & Personal data tracker & 1.5.1 & Using a | symbol in the name of the first option in multiple option values causes the app to crash. \\ 
Track \& Graph & Personal data tracker & 1.5.1 & A wrong number is saved if the used decimal separator does not match the one defined in the Android settings. \\ 
\bottomrule
\end{tabular}
}
\end{table*}

\section{Experiment Design}
\label{sec:experiment-design}

\subsection{Goals and Research Questions}
The goal of this study is to investigate the impact of privacy-preserving techniques on the capability to reproduce the failures experienced in the field. To this end, we framed the following research questions.

\noindent \textbf{\emph{RQ1 - Effectiveness}: What is the failure-reproduction rate for anonymized failure traces?}
\textnormal{This research question studies how failure-reproducing test cases derived from failure traces are impacted by privacy-preserving techniques. That is, it investigates how hard reproducing failures is, if the source traces are anonymized with the techniques presented in Section~\ref{sec:background}.}

\noindent \textbf{\emph{RQ2 - Cost}: How many runs are necessary to reproduce failures with high confidence?}
This research question studies the number of test executions that must be performed to establish if either the failure has been reproduced or the failure cannot be reproduced from the anonymized trace.

\noindent \textbf{\emph{RQ3 - Information Disclosure}: How often is the original value disclosed?}
\textnormal{This research question investigates how often the\linebreak anonymized value is reconstructed while reproducing a failure, thus potentially revealing sensitive information that should remain hidden.}

\subsection{Selection of the Subject Android Apps and Faults} \label{sec:appSelection}

To select the subject apps and faults, we performed an extensive manual analysis to look for failures that depend on user data, that is, failures that can be observed only if certain input values are entered. We restricted our selection to open-source F-Droid~\cite{f-droid} apps with repositories present in either GitHub or GitLab, to make sure it is possible to inspect the app and actually identify the fault responsible for a given failure.
We considered apps in the Money, Science \& Education, Sports \& Health, Time, Internet, and Writing categories, since these apps have a better chance of exploiting user inputs than apps in categories like Theming and Connectivity.

For every category, we manually checked at least 50 apps per category to identify the ones that have fillable fields, considering both the screenshots and the descriptions on their own page in F-Droid. 
For every identified app, we checked all the issues labeled as ``bug'' (or simply all the issues when labels are not available) to select the issues that are reported to be caused by specific inputs. We identified a total of 29 potentially useful issues spanning 26 apps.

To verify the presence of these issues, we downloaded the APK file corresponding to the version with the issue and reproduced the failure as reported in the issue. When the APK was not available, we checked out the correct version from the GitLab or GitHub repository of the application and generated the compiled app ourselves with Android Studio. We also checked the code of the app to determine if two same failures of a same app were originated by a same fault. We then classified failures as \textit{reproducible} or \textit{non-reproducible}, depending on the possibility to reproduce the failure with either an automatic Espresso~\cite{espresso} test case or a failure reproducing routine. In particular, we consider a failure non-reproducible if we could neither reproduce it with Espresso nor we could establish a clear relationship between the inputs and the fault present in the app. %

We found eight non-reproducible failures and two identical failures generated by faults that were already included in the selection. We ended up with \bugs reproduced input-dependent failures caused by distinct bugs in \apps Android apps.
Detailed information about all the considered cases is publicly available in our repository, alongside with the material needed to reproduce the experiments and the results that we obtained:  \repoExperiments. \autoref{tab:app-description} reports the apps, their domain and version, and a description of the bugs present in the apps.%

For each reproduced failure, with the exception of two cases where the app was incompatible with Espresso, we recorded an automatic Espresso test case that exposes it. For the two cases of incompatibility, we inspected the faulty code in the apps and implemented a failure reproducing routine that given an anonymized input determines if the fault is exposed. %

\subsection{Configuration of the Privacy-Preserving Techniques} \label{sec:configurations}

Depending on the nature of the data that must be anonymized, the techniques presented in Section~\ref{sec:background} may require to be properly configured. In the following, we describe the configurations that we used.

\emph{String values} can be anonymized with the Local Suppression and the Special Char Driven Local Suppression techniques. In both cases, the new value that must replace the original one is obtained according to a regular expression. We use the following four regular expressions that capture the cases we encountered in our subject apps: \verb|[!-|\textasciitilde\verb|]|, when all possible string values including special characters are allowed; \verb|[A-Za-z0-9]|, when only alphanumeric values are allowed; \verb|[0-9.,]|, when only numbers with any decimal separator are allowed; and \verb|[0-9,]| or \verb|[0-9.]| or \verb|[0-9:]|, when only number with specific separators are allowed. All these cases are summarized in table \autoref{tab:string-configurations}.   

\begin{table}[ht]
\centering
\caption{Regular expressions for string values.}
\label{tab:string-configurations}
\begin{tabularx}{\columnwidth}{lX}
\toprule
Value type & Regex  \\
\midrule
All possible string values with special char. & \verb|[!-|\textasciitilde\verb|]| \\
Alphanumeric values & \verb|[A-Za-z0-9]| \\
Numbers allowing both the decimal separators & \verb|[0-9.,]| \\
Numbers allowing only a specific separator & \verb|[0-9,]| or \verb|[0-9.]| or \verb|[0-9:]|  \\
\bottomrule
\end{tabularx}
\end{table}

When the Special Char Driven Local Suppression technique is used and the original string includes one or more special characters, these special characters are inserted in random places within the new anonymized string.
We configure the length of the generated string in two ways, experiencing both in our evaluation. That is, the length of the generated string can be random or equal to the length of the original string. In the case of random length, we use the interval [1-25] for short inputs (e.g., a note title or a loyalty card name) and [1-150] for long inputs (e.g., a description). In case the length of an input is bounded to a value lower than the maximum defined by these intervals, we set the maximum length to the maximum length accepted by the text field.

\emph{Numeric values} can be anonymized with most of the privacy-preserving techniques. Local Suppression anonymizes values by generating new values within a specified interval. If the value to anonymize has boundaries defined by the application (e.g., the time can be only assigned with a value in the interval [0-24]), we configure the technique with these limits. Otherwise, we set the interval depending on the nature of the value: when a small value is expected (e.g., an age), we use the interval [0-100], otherwise if bigger values can be used (e.g., a currency) we use the interval [0-1.000.000].
Global Recoding and Rounding require the definition of the number of partitions to be used to split the interval of definition. Consistently with the previous definition of small and big values, we run the techniques with three configurations (using 2, 3, and 4 partitions) when a small value is expected by the app, and we use three different configurations (using 50, 100, and 500 partitions) when a big value is expected.
Finally, we experience three different noise values (30\%, 40\%, and 50\%) for Noise Addition.

The configurations that we used for the techniques applicable to numeric values are summarized in \autoref{tab:number-configurations}.

\begin{table}[ht]
\centering
\caption{Configurations for techniques applied to numbers.}
\label{tab:number-configurations}
\begin{tabularx}{\columnwidth}{XlX}
\toprule
Technique & Parameters & Possible Config.  \\
\midrule
All & Interval of Definition & 1) As in the app \newline 2) [0-100] \newline 3) [0-1.000.000]\\\\
Global Recoding \& Rounding & Number of Partitions & 1) Small intervals: 2, 3, and 4 \newline 2) Big intervals: 50, 100, and 500 \\\\
Noise Addition & Width of Noise & 1) 30\%, 40\%, 50\% \\
\bottomrule
\end{tabularx}
\end{table}

\subsection{Experimental Procedure}

\begin{figure*}[ht]
\begin{center}
    \includegraphics[width=2\columnwidth]{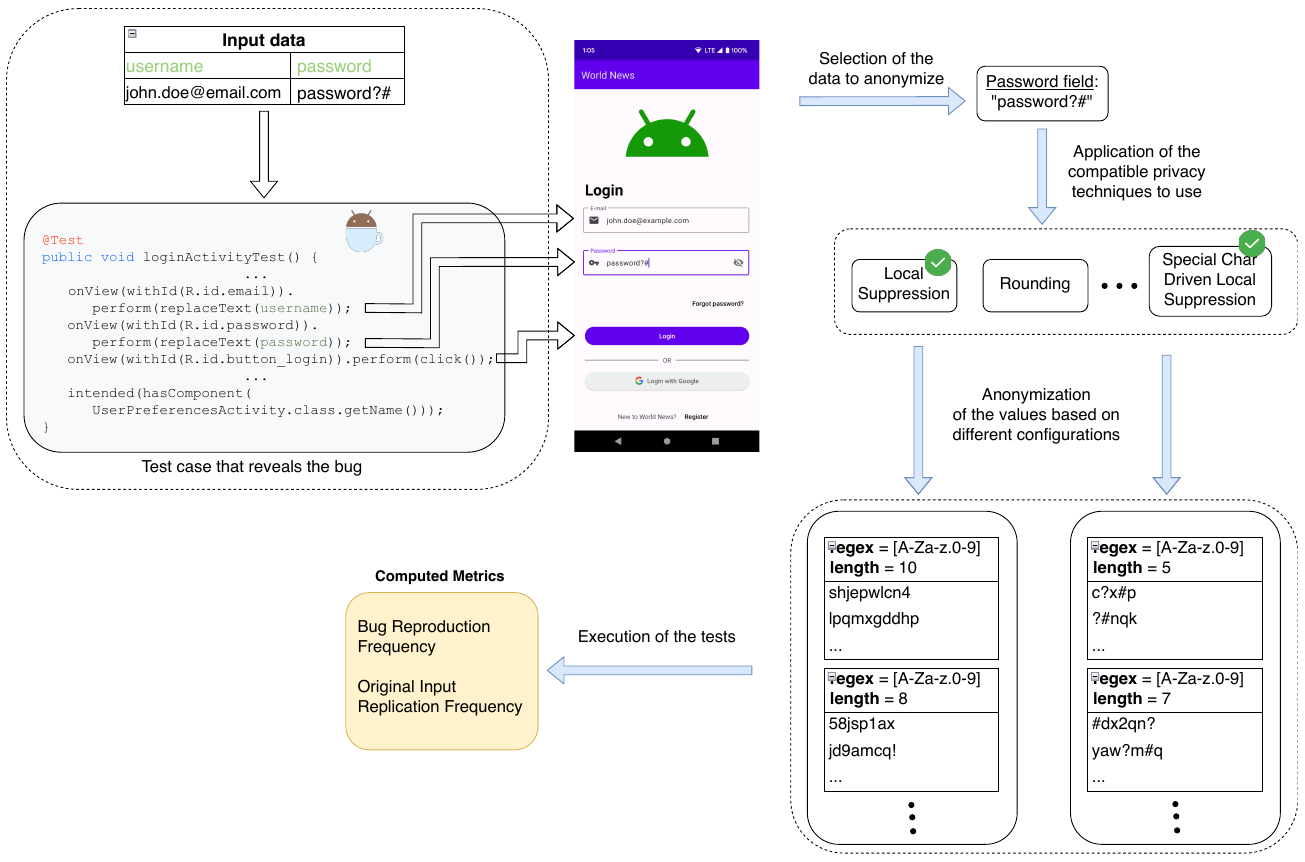}
    \end{center}
    \captionsetup{width=1.8\columnwidth, margin={0\columnwidth, 0\columnwidth}, justification=centering} %
    \captionof{figure}{Overview of the experiment.}
    \label{fig:experiment-overview}
\end{figure*}

To answer RQ1-3, we follow the procedure visually illustrated in \autoref{fig:experiment-overview}. We start from the Espresso test case that reproduces the bug as reported from the field by the user of the application, including the data reported in the original online issue. We identify ourselves a value coherent with the description in the issue, in the few cases a specific value was not available.
To study the impact of privacy-preserving techniques, we anonymize the user data that the failure is dependent on with every applicable technique configured as discussed in Section~\ref{sec:configurations}. %

The anonymization of the data resulted in a new Espresso test case that uses the values derived from the anonymization process rather than the original values. We then executed the new test and checked if the same failure could be reproduced after the anonymization process. Since anonymization and failure reproduction imply randomness, we repeat the anonymization process $100$ times for every configuration, for a total of more than 11K test executions. All tests were executed on a Huawei P9 Lite smartphone with Android 9, except for the few cases that required a specific Android version and were tested on virtual devices. In the two cases of apps incompatible with Espresso, we executed our failure reproducing routine. 

The implementation of the privacy-preserving techniques presented in this paper and the tool to run the failure reproduction process are publicly available in the following repository: \repoTool.

The set of applications used in the study, the input that has been anonymized, the technique used for the anonymization, and the configurations used for the anonymization process are reported in detail in table \autoref{tab:techniques-bugs-config}. We add the labels Lo, Me, Hi next to the configurations present in the table, to identify the configurations that retain less (Lo), medium (Me), or more (Hi) information from the original non-anonymized value.

\input{configurations-table-test}

To answer RQ1, we measure the \emph{bug reproduction frequency}, that is, the ratio between the number of anonymized tests that reveal the same failure revealed by the original test and the number of anonymized tests. The more often a failure is revealed, the less impact a privacy-preserving technique has on the failure reproduction capability of the test cases. To answer RQ2, we compute the \emph{number of repetitions} to reveal the original failure with a probability of 95\%, that is, we estimate the number of tests that must be derived and executed from failure traces to be reasonably sure that a bug has been either reproduced or it is not feasible to reproduce it. To answer RQ3, we measure the \emph{replication frequency of the original input}, that is, we measure the number of times the original non-anonymized value is generated during the failure reproduction process. %

\section{Results}
\label{sec:results}

\subsection{RQ1 - Effectiveness}

\begin{table*}[ht]%
\captionof{table}{Bug Reproduction Frequency}\label{tab:bugReproduction}%
\resizebox{\textwidth}{!}{
\begin{tabular}{lccccclccccccccccc} 
\toprule
& & \multicolumn{4}{c}{Strings} & & &\multicolumn{10}{c}{Numbers} \\
 \cmidrule(lr){3-6} \cmidrule(lr){9-18}
 & & \multicolumn{2}{c}{Local Sup} &  \multicolumn{2}{c}{SCD Local Sup} & & &\multicolumn{1}{c}{Local Sup} &\multicolumn{3}{c}{Global Recoding} &\multicolumn{3}{c}{Rounding} &\multicolumn{3}{c}{Noise Addition} \\
\cmidrule(lr){3-4} \cmidrule(lr){5-6} \cmidrule(lr){9-9} \cmidrule(lr){10-12} \cmidrule(lr){13-15} \cmidrule(lr){16-18}
  &   &  Lo & Hi & Lo & Hi  & & & & Lo & Me & Hi & Lo & Me & Hi & Lo & Me & Hi \\
\midrule
\multicolumn{2}{l}{Binary Eye} & 0\% & 0\% & 0\% & 0\% &  \multicolumn{2}{l}{Birday}  & 0\% & 0\% &2\% & 1\% & 0\% & 0\% &0\% & 6\% & 5\% & 8\% \\
\multicolumn{2}{l}{Catima Loyalty - bug1}  & 8\% & 10.5\% & 9\% & 9.5\% & \multicolumn{2}{l}{Catima Loyalty - bug 2} & 39\% & 62\% & 100\% &39\% & 100\% & 100\% & 0\% &56\% & 64\%  & 60\%\\
\multicolumn{2}{l}{Contact Diary} & 8\% & 16\% & 49\% & 38\% & \multicolumn{2}{l}{Debitum}  & 5\% & 9\% &5\% & 6\% & 0\% & 0\%  &0\% & 6\%  & 5\% & 6\% \\
\multicolumn{2}{l}{EinkBro}  & 0\% & 1\% & 3\% & 7\% & \multicolumn{2}{l}{Did I Take My Meds} & 52\% & 100\% & 49\% & 100\%  & 100\% & 0\%  &100\%   & 74\%  & 84\% & 90\%  \\
\multicolumn{2}{l}{Food Scale Droid}   & 61\% & 28\% & 92\% & 90\% & \multicolumn{2}{l}{Grow Tracker} & 100\% & 100\% &100\%   & 100\%  & 100\% & 100\% & 100\%   & 100\%  & 100\% & 100\%  \\
\multicolumn{2}{l}{Grow Tracker} & 77\% & 27\% & 100\% & 100\% & \multicolumn{2}{l}{Money Wallet} & 4\% & 5\% & 9\% & 13\%  & 0\% & 0\% & 0\% & 8\%  & 8\% & 8\% \\
\multicolumn{2}{l}{NoNonsense Notes}   & 18\% & 8\% & 100\% & 100\% & \multicolumn{2}{l}{Simple Calendar} & 27\% & 74\% & 100\% & 36\% & 100\%  & 100\% & 0\% & 63\% & 63\%  & 61\% \\
\multicolumn{2}{l}{Simple Money Tracker}   & 1\% & 16\% & 5\% & 15\% & \multicolumn{2}{l}{Track \& Graph - bug 2} & 100\% & 100\% &100\%   & 100\%  & 100\% & 100\% & 100\%   & 100\%  & 100\% & 100\%  \\
\multicolumn{2}{l}{SplitBills}  & 10\% & 9\% & 81\% & 93\%  \\
\multicolumn{2}{l}{Tasks}  & 0\% & 0\% & 20\% & 6\% &  \\
\multicolumn{2}{l}{To Don't} & 18\% & 9\% & 99\% & 99\% &  \\
\multicolumn{2}{l}{Track \& Graph - bug 1}   & 3\% & 2\% & 13\% & 21\% &  \\
\multicolumn{2}{l}{Track \& Graph - bug 2}   & 35\% & 15\% & 59\% & 53\% & \\[0.15cm]
\textbf{Mean} & &  \textbf{18\%} & \textbf{11\%} & \textbf{49\%} & \textbf{49\%} & & & \textbf{41\%} & \textbf{56\%} & \textbf{58\%} & \textbf{49\%} & \textbf{63\%} & \textbf{50\%} & \textbf{38\%} & \textbf{52\%} & \textbf{54\%} & \textbf{54\%} \\
\bottomrule
\multicolumn{18}{l}{\textit{'Lo', 'Me' and 'Hi' in the header refer to configurations that retain less, medium, or more information from the input}}
\end{tabular}

}

\end{table*}

\autoref{tab:bugReproduction} reports the bug reproduction frequency for the privacy-preserving techniques applied to strings and numbers. 

Concerning the anonymization of strings, Local Suppression severely compromises the capability to reproduce failures (the mean success rate varies between 11\% and 19\%), depending on the configuration. The low bug reproduction frequency for Local Suppression is expected, since almost no information is retained from the original string. Interestingly, retaining more information from the original input (configuration Hi) has, in the majority of the cases, a negligible or negative effect on the bug reproduction frequency. This happens because preserving the length of the original string is often not a relevant factor in failure reproduction, while using (longer) random strings may increase the chance of using the right combination of characters that may trigger the failure.  

In line with this intuition, SCD Local Suppression has a significantly better performance than Local Suppression (mean success rate of 49\%). This confirms our intuition that by just disclosing a syntactic information that is largely irrelevant on the point of view of the user (i.e., the presence of a special character), failure reproduction might be often improved. Again, retaining the length of the string has not an impact on failure reproduction.
Clearly, the presence of special characters alone is not always enough to reproduce failures. In these cases Local Suppression and SCD Local Suppression have similar performance, as for the Catima Loyalty and the Binary Eye apps. In some other cases, they are helpful but not sufficient alone, since the special character(s) might have to occur at a specific position, as in the first bug of the Track \& Graph app, or in the context of a specific string, as in the Task app. %
In some other cases the fault was dependent on the semantic of the value and the mere presence of the special character was not enough to reproduce the failure.

For numeric inputs, Local Suppression is the technique with the lowest success rate (41\%), with only Rounding - Hi performing worst (38\%). Noise Addition and Global Recoding perform similarly: the effectiveness of Global Recoding ranged between 49\% and 58\%, and Noise Addition ranged between 52\% and 54\%. Rounding performed best in some cases, but with higher performance variance, with an effectiveness between 38\% and 63\%.  
Global Recoding and Rounding are more sensitive than Noise Addition to the choice of the configuration. In fact, Noise Addition works with intervals that are defined around the original input value. On the contrary, the intervals used in Global Recoding and Rounding are independent from the original input value, which could fall very near or on the edge of the interval, affecting the reproduction probability in cases where the fault is caused by values near to the original one.

\begin{table*}[ht]
\captionof{table}{Iterations for Reproducing Failures with 95\% probability}\label{tab:bugcost}%
\resizebox{\textwidth}{!}{
\begin{tabular}{lccccclccccccccccc} 
\toprule
& & \multicolumn{4}{c}{Strings} & & &\multicolumn{10}{c}{Numbers} \\
 \cmidrule(lr){3-6} \cmidrule(lr){9-18}
 & & \multicolumn{2}{c}{Local Suppression} &  \multicolumn{2}{c}{SCD Local Suppression} & & &\multicolumn{1}{c}{Local Sup} &\multicolumn{3}{c}{Global Recoding} &\multicolumn{3}{c}{Rounding} &\multicolumn{3}{c}{Noise Addition} \\
\cmidrule(lr){3-4} \cmidrule(lr){5-6} \cmidrule(lr){9-9} \cmidrule(lr){10-12} \cmidrule(lr){13-15} \cmidrule(lr){16-18}
  &   &  Lo & Hi & Lo & Hi  & & & & Lo & Me & Hi & Lo & Me & Hi & Lo & Me & Hi \\
\midrule
\multicolumn{2}{l}{Binary Eye} & - & - & - & - &  \multicolumn{2}{l}{Birday}  & - & - & 149 & 299 & - & - & - & 49 & 59 & 36 \\
\multicolumn{2}{l}{Catima Loyalty - bug 1}  & 36 & 28 & 32 & 31 & \multicolumn{2}{l}{Catima Loyalty - bug 2} & 7 & 4 & 1 & 7 & 1 & 1 & - & 4 & 3  & 4\\
\multicolumn{2}{l}{Contact Diary} & 36 & 18 & 5 & 7 & \multicolumn{2}{l}{Debitum}  & 59 & 32 & 59 & 49 & - & -  & - & 49 & 59 & 49 \\
\multicolumn{2}{l}{Einkbro}  & - & 299 & 99 & 42 & \multicolumn{2}{l}{Did I Take My Meds} & 5 & 1 & 5 & 1  & 1 & -  &1   & 3  & 2 & 2  \\
\multicolumn{2}{l}{Food Scale Droid} & 4 & 10 & 2 & 2 & \multicolumn{2}{l}{Grow Tracker} & 1 & 1 & 1 & 1 & 1 & 1 & 1 & 1  & 1 & 1  \\
\multicolumn{2}{l}{Grow Tracker} & 3 & 10 & 1 & 1 & \multicolumn{2}{l}{Money Wallet} & 74 & 59 & 32 & 22 & - & - & - & 36  & 36 & 36 \\
\multicolumn{2}{l}{NoNonsense Notes}   & 16 & 36 & 1 & 1 & \multicolumn{2}{l}{Simple Calendar} & 10 & 3 & 1 & 7 & 1  & 1 & - & 4 & 4  & 4 \\
\multicolumn{2}{l}{Simple Money Tracker}   & 299 & 18 & 59 & 19 & \multicolumn{2}{l}{Track \& Graph - bug 2} & 1 & 1 & 1 & 1 & 1 & 1 & 1 & 1  & 1 & 1  \\
\multicolumn{2}{l}{SplitBills}   & 29 & 32 & 2 & 2 & \\
\multicolumn{2}{l}{Tasks}  & - & - & 14 & 49 &  \\
\multicolumn{2}{l}{To Don't} & 16 & 32 & 1 & 1 &  \\
\multicolumn{2}{l}{Track \& Graph - bug 1}   & 99 & 149 & 22 & 13 &  \\
\multicolumn{2}{l}{Track \& Graph - bug 2}   & 7 & 19 & 4 & 4 & \\[0.15cm]
\textbf{Mean*} & & \textbf{55} & \textbf{60} & \textbf{21} & \textbf{15} & & & \textbf{23} & \textbf{15} & \textbf{32} & \textbf{49} & \textbf{1} & \textbf{1} & \textbf{1} & \textbf{19} & \textbf{21} & \textbf{17} \\
\textbf{Max} & & \textbf{299} & \textbf{299} & \textbf{99} & \textbf{49} & & &\textbf{74} & \textbf{59} & \textbf{149} & \textbf{299} & \textbf{1} & \textbf{1} & \textbf{1} & \textbf{49} & \textbf{59} & \textbf{49}\\
\textbf{\# NR}    &   & \textbf{3} & \textbf{2} &\textbf{1} & \textbf{1} & & &\textbf{1} &  \textbf{1} & \textbf{0} & \textbf{0} & \textbf{3} & \textbf{4} & \textbf{5} & \textbf{0} & \textbf{0} & \textbf{0}\\
\bottomrule
\multicolumn{18}{l}{\textit{'Lo', 'Me' and 'Hi' in the header refer to configurations that retain less, medium, or more information from the input} } \\
\multicolumn{18}{l}{\textit{*mean value is calculated only over defined values (- cases are ignored in the computation of the mean)}}
\end{tabular}}
\vspace{-0.3cm}
\end{table*}

The characteristics of the failure to be reproduced also have an impact. In fact, there are some easy cases where most of the techniques were systematically successful, for instance due to values formatted according to the Android settings %
that systematically generate failures if incompatible with the expectation of the app. We also had some hard cases with a failure reproduction rate below 10\%. This is due to the small set of domain values that trigger the failure (e.g., in Birday only February 29 of leap years lead to the malfunction, over all the possible dates). In cases where the bug was caused by values in a range close to the original input (e.g., Simple Calendar, Did I Take My Meds, Catima Loyalty), Local Suppression is affected by its inability to preserve any  information, leading to an overall lower success rate compared to other techniques. Noticeable, although if with low probability, it was always possible to replicate a bug using Global Recoding and Noise Addition, while Rounding tends to quickly reveal the failure or miss it.

For three apps, the attempt to reproduce the original failure led to the discovery of new bugs. In Binary Eye some strings used to generate a QR code differ from the ones obtained when scanning the code generated by the app (e.g., \texttt{\}c:+8ha} when coded and then decoded becomes \texttt{$\sim$c:+8ha}). In To Don't some task names when saved cause all the other task names to be cancelled and replaced by a null value (e.g., the random string \texttt{S\,0O\}\_('} was sufficient to reveal the bug). In Track \& Graph - bug 1, the bug makes the app crash while creating a new multiple-value tracked habit, when the first option ends with `$\mid$', but we also discovered that option names with `$\mid$$\mid$' cause the habit to not be saved.

\textbf{Answer to RQ1:}
SCD Local Suppression should be preferred to Local Suppression when applied to strings, since it significantly increases the failure reproduction capability, while disclosing minimal information (the presence of a special character in the original string). 
Local Suppression applied to numbers had a significant, but not dramatic, impact on failure reproduction (41\% success rate). Techniques approximating and perturbing the original value might increase the success rate (up to 63\% in our experiments) at the cost of disclosing some information about the original value.
Deciding how much information preserving from the input values should be done carefully, since preserving information not correlated with the failure trigger may negatively influence reproduction.

\subsection{RQ2 - Cost}

To measure the cost of using anonymized data to reproduce failures instead of using actual values, we computed the number of attempts (i.e., generation of anonymized values and then generation of the concrete test cases) that must be completed to reproduce the original failure with a probability of 95\%. \autoref{tab:bugcost} shows the Mean and Max number of attempts necessary across all faults for a given technique and configuration, the number of non-reproduced faults (row \# NR), and the results per fault. Note that a higher average success rate does not imply fewer attempts in average since there is a logarithmic relationship between reproduction probabilities and number of attempts.

When anonymizing strings, Local Suppression introduces a cost hardly affordable in practice, with close to 60 test generations and executions attempts needed in average, and up to 299 attempts in the worst case. SCD Local Suppression is more practical since it requires between 15 and 21 attempts in average, with a maximum between 49 and 99.

When working with numbers, Noise Addition seems to be the most affordable solution, since it reproduced all failures with 95\% confidence with a mean number of attempts between 17 and 21 and up to 59 in the worst case (Me). Global Recoding is a good choice too, as it performs similar to Noise Addition except with Birday, which determines the higher mean and max values for this technique. This difference is due to the necessity of preserving day and month unchanged in the original input date (29-02-1996) to reproduce the failure, which is more likely to happen with Noise Addition since it creates intervals around the input value. Rounding can be useful to reduce the failure reproduction effort, but it also severally reduces the number of reproduced failures. Local Suppression is a valid alternative to Noise Addition when no information about the original value has to be disclosed, at the risk of failing to reproduce some failures.

\textbf{\textbf{Answer to RQ2:}}
SCD Local Suppression is a cost-effective solution to anonymize strings. Numbers can be feasibly addressed with Noise Addition or Local Suppression, depending on the amount of information %
that can be disclosed.

\subsection{RQ3 - Information Disclosure}
\begin{table}[ht]
\captionof{table}{Frequency of replication of the original value.}\label{tab:frequency} \vspace{-0.3cm}
\resizebox{0.9\columnwidth}{!}{
\begin{tabular}{lll} 
\toprule
 App    &  Technique & Replication Freq. \\
\midrule
\multicolumn{3}{c}{\emph{String}}\\
Track \& Graph - 2 & Local Suppression & 0.50\%\\

\multicolumn{3}{c}{\emph{Numbers}}\\

Birday & Global Recoding & 0.67\% \\
Birday & Noise Addition & 1.67\% \\
Catima Loyalty - 2 & Noise Addition &0.67\%\\
Debitum &Noise Addition & 0.33\%\\

\bottomrule

\end{tabular}}
\end{table}

We computed how often the failure reproduction process has generated the value before the anonymization during failure reproduction. \autoref{tab:frequency} shows the percentage of cases it happened for the various combinations of apps and techniques. 
In case of strings, obtaining the original value is unlikely to happen due to the size of the space of possibilities. In fact, it happened only once for one app, where the format of the string was particularly constrained by the regular expression. 
In case of numbers, the replication of the original value happened slightly more frequently, with Noise Addition being responsible of the highest number of cases (which anyway consists of only three cases with a probability below 1.67\%). This is due to the relatively smaller size of the numeric domain and the type of perturbations introduced by Noise Addition.
Interestingly, it was not strictly necessary to reconstruct the original value in any of these cases, so the reproduction of the value was incidental and the user would not be really aware of this fact. The only exception is Birday where the failure requires the date \texttt{29-2-year} where \texttt{year} is any leap year. Thus the user would discover the date and month of the birthday, and would restrict the birthday to leap years.

\textbf{\textbf{Answer to RQ3:}}
All the anonymization techniques largely hide the values they are applied to. Sometime the failure reproduction process may generate the original input in the attempt to reproduce the failure. This is unlikely to happen frequently, with only Noise Addition causing the reproduction of the original input in some rare cases.

\subsection{Threats to Validity}
\label{sec:threats-to-validity}

A threat is about the limited set of bugs considered in the experiments. To mitigate this threat, we systematically searched for real bugs contained in open source applications and we selected Android applications from different categories to have multiple contexts in which to experiment with privacy-preserving techniques. The construction of the experimental dataset that we publicly released is already the result of significant manual effort with hundreds of apps and reports manually inspected, as described in Section~\ref{sec:appSelection}. Enlarging this dataset to address new domains is part of our future work.

Another concern is about the way we configured the privacy-preserving techniques. To avoid introducing any bias, we defined a configuration policy that we described in the paper. All the configurations are finally reported in our online repository. %

Finally, another concern is related to the correctness of the implementation of the privacy-preserving techniques that we used for the experiments. To mitigate this threat, we extensively tested our tools and made our artifacts publicly available.

\section{Related Work}
\label{sec:related-work}

The studies most related to our work concern with the approaches designed for the anonymization of the data collected during failures, and with solutions for bugs reproduction.

One of the first approaches designed for releasing private data in the context of testing and debugging activities, while ensuring people's privacy, is \textit{kb}-Anonymity~\cite{DBLP:conf/pldi/BudiLJL11}. This approach exploits symbolic execution and \textit{k}-anonymity to generate anonymized database tuples that do not alter the behavior of the program, that is, the same program path is executed when the program uses  both the original and anonymized values. The approach is limited to numbers and programs whose code is accessible and analyzable with symbolic execution. Castro et al.~\cite{DBLP:conf/asplos/CastroCM08} investigated a similar approach but applied to the data included in crash reports. %
MultiPathPrivacy~\cite{DBLP:conf/edcc/LouroGR12} and RESPA~\cite{DBLP:conf/issre/MatosGR15} investigated how to weaken the requirement about preserving the same execution path when introducing anonymized values by identifying alternative paths that shall still lead to the reproduction of the same bug. 

Different from this body of work, we studied the effectiveness of privacy-preserving techniques that have been extensively applied to databases and that can be easily used to anonymize data collected during failures, without running any complicated analysis on the code of the application. The results reported in this paper provide useful insights about their effectiveness and cost, and the specific configurations that best fit the problem of anonymizing failure data.

Our work also relates to failure reproduction. We target the case of reproducing failures from (anonymized) failure traces collected from Android applications. The reproduction of failures from similar \emph{non-anonymized} traces has been also considered in other works, such as CaRCrash~\cite{DBLP:conf/qrs/SunYLZX19} that collects and dispatches failures traces every time a failure is detected. Similarly, CrashDroid~\cite{DBLP:conf/iwpc/WhiteVJBP15} can reconstruct replayable scripts from stack traces collected during failures.

Some other techniques considered reproducing failures from bug reports using NLP techniques, such as  S2RMiner~\cite{DBLP:conf/icsr/ZhaoMYZP19} and ReCDroid~\cite{DBLP:conf/icse/ZhaoYSLZZH19}. In this study, we considered the impact of anonymization techniques on failure traces and the corresponding test cases. We left to future work investigating  more in details the impact of privacy-preserving techniques on test cases derived from bug reports, although in principle the artefact used to derive the test cases should not significantly affect the conclusions of our study.

Finally, some techniques addressed the problem of reproducing failures in Java, such as, BugRedux~\cite{DBLP:conf/icse/JinO12}, JCHARMING~\cite{DBLP:conf/wcre/NayrollesHTL15}, STAR~\cite{DBLP:journals/tse/ChenK15}, and EvoCrash~\cite{DBLP:journals/tse/SoltaniPD20}. We targeted Android since apps are often used to process personal information. Investigating other technical contexts is part of our future work.

\section{Conclusions}
\label{sec:conclusion}

Analyzing and reproducing failures from failure traces is important to timely fix faults and develop reliable applications. %
However, failure traces may disclose sensitive information about the users of the applications, and must be properly anonymized before they can be used for failure reproduction.
This paper studies how privacy-preserving techniques extensively exploited in the context of database systems can be adapted to the problem of failure reproduction, and presents an empirical evaluation that discloses findings about their effectiveness and cost. In particular, our results show that the SCD Local Suppression technique introduced in this paper can be effective with the anonymization of strings, while numbers can be effectively anonymized with Local Suppression or Noise Addition, depending on the possibility to disclose some information about the original value that was anonymized.
Our future work concerns with experiencing and studying privacy-preserving techniques applied to additional domains, such as Web Applications. %

\begin{acks}
This work has been partially supported by the Engineered MachinE Learning-intensive IoT systems (EMELIOT) national research project (PRIN 2020 program Contract 2020W3A5FY).
\end{acks}

\bibliographystyle{ACM-Reference-Format}
\bibliography{IEEEabrv,bibliography}

\end{document}

%% file: configurations-table-test.tex
\begin{table*}
\centering
\caption{Configurations of the techniques for each application's bugs analyzed.}
\label{tab:techniques-bugs-config}
\scriptsize
\resizebox{2\columnwidth}{!}{%
\begin{tabular}{llll}
\hline
App bug & Input & Technique & Configuration \\

\hline

\multirow{3}{*}{Binary Eye} & \multirow{3}{*}{\shortstack[l]{com.taobao.arthas.boot\\.ProcessUtils.findJavaHome\\(ProcessUtils.java:222)}}  & \multirow{3}{*}{\shortstack[l]{Local Suppression \&\\ SCD Local Suppression}} & \multirow{3}{*}{\textbf{regex}: [!- $\sim$], \textbf{length}: equal to original (Hi) or [1-150] (Lo)} \\\\\\

\hline

 \multirow{5}{*}{Birday} & \multirow{5}{*}{29 2 1996} & Local Suppression & \textbf{interval}: [1-31] [1-12] [1937-2036] \\

\addlinespace[1.3ex] & & \multirow{2}{*}{\shortstack[l]{Global Recoding \&\\ Rounding}} & \textbf{interval}: [1-31] [1-12] [1937-2036], \textbf{partitions}: 2 (Lo) or 3 (Me) or 4 (Hi) \\\\

\addlinespace[1.3ex] & & \multirow{1}{*}{Noise Addition} & \textbf{interval}: [1-31] [1-12] [1937-2036], \textbf{noise width}: 0.3 (Hi) or 0.4 (Me) or 0.5 (Lo) \\

\hline

\multirow{2}{*}{Catima Loyalty - bug 1} & \multirow{2}{*}{Atelier} & \multirow{2}{*}{\shortstack[l]{Local Suppression \&\\ SCD Local Suppression}} & \textbf{regex}: [!- $\sim$], \textbf{length}: equal to original (Hi) or [1-25] (Lo) \\
& & & \textbf{regex}: [A-Za-z0-9 ], \textbf{length}: equal to original (Hi) or [1-25] (Lo)\\

\hline

\multirow{5}{*}{Catima Loyalty - bug 2} & \multirow{5}{*}{19 4 1963} & Local Suppression &  \textbf{interval}: [1-31] [1-12] [1900-2100] \\

\addlinespace[1.3ex] & & \multirow{2}{*}{\shortstack[l]{Global Recoding \&\\ Rounding}} & \textbf{interval}: [1-31] [1-12] [1900-2100], \textbf{partitions}: 2 (Lo) or 3 (Me) or 4 (Hi) \\\\

\addlinespace[1.3ex] & & \multirow{1}{*}{Noise Addition} &  \textbf{interval}: [1-31] [1-12] [1900-2100], \textbf{noise width}: 0.3 (Hi) or 0.4 (Me) or 0.5 (Lo) \\

\hline

\multirow{2}{*}{Contact Diary} & \multirow{2}{*}{:30}  & \multirow{2}{*}{\shortstack[l]{Local Suppression \&\\ SCD Local Suppression}} & \multirow{2}{*} {\textbf{regex}: [0-9:], \textbf{length}: equal to original (Hi) or [1-5] (Lo)} \\\\

\hline

\multirow{5}{*}{Debitum} & \multirow{5}{*}{4.60} & Local Suppression & \textbf{interval}: [0-100) \\

\addlinespace[1.3ex] & & \multirow{2}{*}{\shortstack[l]{Global Recoding \&\\ Rounding}} & \textbf{interval}: [0-100), \textbf{partitions}: 2 (Lo) or 3 (Me) or 4 (Hi) \\\\

\addlinespace[1.3ex] & & \multirow{1}{*}{Noise Addition} & \textbf{interval}: [0-100), \textbf{noise width}: 0.3 (Hi) or 0.4 (Me) or 0.5 (Lo) \\

\hline

\multirow{5}{*}{Did I Take My Meds} & \multirow{5}{*}{15:30 20:36} & Local Suppression & \textbf{interval}: [0-23] [0-59] [0-23] [0-59] \\

\addlinespace[1.3ex] & & \multirow{2}{*}{\shortstack[l]{Global Recoding \&\\ Rounding}} & \textbf{interval}: [0-23] [0-59], \textbf{partitions}: 2 (Lo) or 3 (Me) or 4 (Hi) \\\\

\addlinespace[1.3ex] & & \multirow{1}{*}{Noise Addition} & \textbf{interval}: [0-23] [0-59], \textbf{noise width}: 0.3 (Hi) or 0.4 (Me) or 0.5 (Lo)\\

\hline

\multirow{2}{*}{EinkBro} & \multirow{2}{*}{how to open design.psd} & \multirow{2}{*}{\shortstack[l]{Local Suppression \&\\ SCD Local Suppression}} & \multirow{2}{*} {\textbf{regex}: [!- $\sim$], \textbf{length}: equal to original (Hi) or [1-25] (Lo)} \\\\

\hline

\multirow{2}{*}{Food Scale Droid} & \multirow{2}{*}{543,} & \multirow{2}{*}{\shortstack[l]{Local Suppression \&\\ SCD Local Suppression}} &  \multirow{2}{*} {\textbf{regex}: [0-9,.], \textbf{length}: equal to original (Hi) or [1-25] (Lo)} \\\\

\hline

\multirow{8}{*}{Grow Tracker} & \multirow{8}{*}{3.6} & \multirow{2}{*}{\shortstack[l]{Local Suppression \&\\ SCD Local Suppression}} & \textbf{regex}: [0-9,.], \textbf{length}: equal to original (Hi) or [1-25] (Lo)\\\\

\addlinespace[1.3ex] & & \multirow{1}{*}{Local Suppression} & \textbf{interval}: [0-100) \\

\addlinespace[1.3ex] & & \multirow{2}{*}{\shortstack[l]{Global Recoding \&\\ Rounding}} & \textbf{interval}: [0-100), \textbf{partitions}: 2 (Lo) or 3 (Me) or 4 (Hi)\\\\

\addlinespace[1.3ex] & & \multirow{1}{*}{Noise Addition} & \textbf{interval}: [0-100), \textbf{noise width}: 0.3 (Hi) or 0.4 (Me) or 0.5 (Lo) \\

\hline

\multirow{5}{*}{Money Wallet} & \multirow{5}{*}{4362.65} & Local Suppression & \textbf{interval}: [0-1000000) \\

\addlinespace[1.3ex] & & \multirow{2}{*}{\shortstack[l]{Global Recoding \&\\ Rounding}} & \textbf{interval}: [0-1000000), \textbf{partitions}: 50 (Lo) or 100 (Me) or 500 (Hi) \\\\

\addlinespace[1.3ex] & & \multirow{1}{*}{Noise Addition} & \textbf{interval}: [0-1000000), \textbf{noise width}: 0.3 (Hi) or 0.4 (Me) or 0.5 (Lo) \\

\hline

\multirow{2}{*}{NoNonsense Notes} & \multirow{2}{*}{list/name} & \multirow{2}{*}{\shortstack[l]{Local Suppression \&\\ SCD Local Suppression}} & \multirow{2}{*} {\textbf{regex}: [0-9,.], \textbf{length}: equal to original (Hi) or [1-25] (Lo)} \\\\

\hline

\multirow{5}{*}{Simple Calendar} & \multirow{5}{*}{1 1 1960} & Local Suppression & \textbf{interval}: [1-31] [1-12] [1900-2100] \\

\addlinespace[1.3ex] & & \multirow{2}{*}{\shortstack[l]{Global Recoding \&\\ Rounding}} & \textbf{interval}: [1-31] [1-12] [1900-2100], \textbf{partitions}: 2 (Lo) or 3 (Me) or 4 (Hi) \\\\

\addlinespace[1.3ex] & & \multirow{1}{*}{Noise Addition} & \textbf{interval}: [1-31] [1-12] [1900-2100], \textbf{noise width}: 0.3 (Hi) or 0.4 (Me) or 0.5 (Lo) \\

\hline

\multirow{2}{*}{Simple Money Tracker} & \multirow{2}{*}{20000000000000000000}  & \multirow{2}{*}{\shortstack[l]{Local Suppression \&\\ SCD Local Suppression}} & \multirow{2}{*} {\textbf{regex}: [0-9.], \textbf{length}: equal to original (Hi) or [1-25] (Lo)} \\\\

\hline

\multirow{2}{*}{SplitBills} & \multirow{2}{*}{group/name}  & \multirow{2}{*}{\shortstack[l]{Local Suppression \&\\ SCD Local Suppression}} & \multirow{2}{*} {\textbf{regex}: [!- $\sim$], \textbf{length}: equal to original (Hi) or [1-25] (Lo)} \\\\

\hline

\multirow{2}{*}{Tasks} & \multirow{2}{*}{Subtask 1 @home} & \multirow{2}{*}{\shortstack[l]{Local Suppression \&\\ SCD Local Suppression}} & \multirow{2}{*} {\textbf{regex}: [!- $\sim$], \textbf{length}: equal to original (Hi) or [1-25] (Lo)} \\\\

\hline

\multirow{2}{*}{To Don't} & \multirow{2}{*}{ task'name    add}  & \multirow{2}{*}{\shortstack[l]{Local Suppression \&\\ SCD Local Suppression}}  & \multirow{2}{*} {\textbf{regex}: [!- $\sim$], \textbf{length}: equal to original (Hi) or [1-25] (Lo)} \\\\

\hline

\multirow{2}{*}{Track \& Graph - bug 1} & \multirow{2}{*}{option$|$}  & \multirow{2}{*}{\shortstack[l]{Local Suppression \&\\ SCD Local Suppression}} & \multirow{2}{*} {\textbf{regex}: [!- $\sim$], \textbf{length}: equal to original (Hi) or [1-25] (Lo)} \\\\

\hline

\multirow{7}{*}{Track \& Graph - bug 2} & \multirow{7}{*}{2.7} & \multirow{2}{*}{\shortstack[l]{Local Suppression \&\\ SCD Local Suppression}} & \textbf{regex}: [!- $\sim$], \textbf{length}: equal to original (Hi) or [1-25] (Lo) \\\\

\addlinespace[1.3ex] & & \multirow{1}{*}{Local Suppression} & \textbf{interval}: [0-100) \\
\addlinespace[1.3ex] & & \multirow{2}{*}{\shortstack[l]{Global Recoding \&\\ Rounding}} & \textbf{interval}: [0-100), \textbf{partitions}: 2 (Lo) or 3 (Me) or 4 (Hi) \\\\

\addlinespace[1.3ex] & & \multirow{1}{*}{Noise Addition} & \textbf{interval}: [0-100), \textbf{noise width}: 0.3 (Hi) or 0.4 (Me) or 0.5 (Lo) \\

\hline
\end{tabular}
}
\end{table*}